\def\year{2021}\relax
\documentclass[letterpaper]{article} 
\usepackage{aaai21}  
\usepackage{times}  
\usepackage{helvet} 
\usepackage{courier}  
\usepackage[hyphens]{url}  
\usepackage{graphicx} 
\usepackage{natbib}  
\usepackage{caption} 

\usepackage[group-separator={,}]{siunitx}

\usepackage{natbib}
\usepackage{amsmath,amssymb,amsfonts}
\usepackage{graphicx}
\usepackage{subcaption}
\graphicspath{ {/plots/without_edge_filtering/} }
\usepackage{url}
\usepackage{xcolor}
\usepackage{dirtytalk}
\usepackage{enumitem}
\usepackage{floatrow}
\usepackage{todonotes}
\usepackage{longtable}

\usepackage{pdfcomment}
\pdfcommentsetup{draft} 
\pdfcommentsetup{color={1.0 1.0 0.0}, open=true}

\title{Identification of Twitter Bots Based on an Explainable Machine Learning Framework: The US 2020 Elections Case Study}

\author {
    Alexander Shevtsov,\textsuperscript{\rm 1,\rm 3}
    Christos Tzagkarakis, \textsuperscript{\rm 1}
    Despoina Antonakaki, \textsuperscript{\rm 1}
    Sotiris Ioannidis, \textsuperscript{\rm 2, \rm 1}\\
}
\affiliations {
    \textsuperscript{\rm 1} Institute of Computer Science, Foundation for Research and Technology\\
    \textsuperscript{\rm 2} Technical University of Crete\\
    \textsuperscript{\rm 3} Computer Science Department - University of Crete\\
    \{shevtsov,tzagarak,despoina\}@ics.forth.gr,  sotiris@ece.tuc.gr
}
\begin{document}

\maketitle
\begin{abstract}
Twitter is one of the most popular social networks attracting millions
of users, while a considerable proportion of online discourse is
captured. It provides a simple usage framework with short messages and
an efficient application programming interface (API) enabling the
research community to study and analyze several aspects of this social
network. However, the Twitter usage simplicity can lead to malicious
handling by various bots. The malicious handling phenomenon expands in
online discourse, especially during the electoral periods, where
except the legitimate bots used for dissemination and communication
purposes, the goal is to manipulate the public opinion and the
electorate towards a certain direction, specific ideology, or
political party.
This paper focuses on the design of a novel system for identifying
Twitter bots based on labeled Twitter data. To this end,
a supervised machine learning (ML) framework is adopted using an Extreme Gradient Boosting (XGBoost) algorithm, where the hyper-parameters are tuned via cross-validation.
Our study also deploys Shapley Additive Explanations (SHAP) for explaining the ML model predictions by calculating feature importance, using the game theoretic-based Shapley values.
Experimental evaluation on distinct Twitter datasets demonstrate the superiority of our approach, in terms of bot detection accuracy, when compared against a recent state-of-the-art Twitter bot detection method.





\end{abstract}

\section{Introduction}
Twitter is considered one of the most popular and widespread online 
social networks (OSNs) nowadays. It is used by millions of users and 
organizations to quickly share and discover information about a service, product, sports/social/political event etc. However, Twitter can be used as an intermediate system for malicious purposes, such as spreading fake news \cite{Bovet2009, sharma2019combating} or manipulating public opinion \cite{badawy2018analyzing}.

Specifically, Twitter can be used to circulate 
propaganda~\cite{neudert2017junk, jones2019gulf,chatfield2015tweeting}, 
manipulate the public opinion~\cite{bolsover2019chinese, seo2014visual}, and influence the electorate towards a particular ideology or political party~\cite{golovchenko2020cross, howard2016bots}. 
These tasks can be fully automated through a special organized group of agents, called \emph{botnets}, which are groups of \emph{sybil} accounts that collectively seek to influence ordinary users. In particular, a botnet is a group of bots, i.e., automated programs programmed to run certain tasks. A \emph{sybil} account in OSNs is a fake identity, not necessarily representing a real person or created by the real person it represents (impersonation technique)~\cite{alsaleh2014tsd}.

It has been observed that Twitter bots can also be exploited to spread fake news, rumors and hate speech~\cite{DBLP:journals/corr/abs-1802-00393,fortuna2018survey, burnap2015cyber} by instantly republishing low credibility Twitter content~\cite{shao2018spread} via popular users and Twitter mentions~\cite{stella2018bots}.

In this work, we aim to build 
a machine learning (ML) framework over a large collected dataset, to detect bot Twitter accounts. We identify and analyze Twitter bots during the US 2020 Elections period. 
The current study provides answers to the following questions:
\begin{itemize}
\item Is it possible to implement and fine-tune a ML-based bot detection model to efficiently apply it to the US 2020 Elections dataset?
\item Which types of features can be extracted from the Twitter application programming interface (API) to promote high performance?
\item Is it possible to examine the ML model's generalization capability in terms of bot detection accuracy across several well-established datasets?
\item Does the proposed ML model act as a black box or could the ML model's mechanism be ``unlocked'' in order to investigate how it yields its predictions?

\end{itemize}
Our analysis can help the research community to better understand the bot detection task and how it can be performed in different types of datasets, or within diverse domains. The presented methodology achieves a high bot detection accuracy on the US 2020 Elections dataset, while attaining increased generalization performance in terms of bot identification when applied on additional, well-established Twitter datasets. The ML model's outcome is also explained based on Shapley Additive Explanations (SHAP) method.

The rest of the paper is organized as follows: \textit{\nameref{sec:background}} explores related past works. A detailed description of the data collection process and the proposed methodology is given in~\textit{\nameref{sec:methodology}}. \textit{\nameref{sec:results}} evaluates the performance of our method, whilst~\textit{\nameref{sec:conclusion}} summarizes the main outcomes and provides directions for future work.

\section{Background}\label{sec:background}

Twitter (social) bots can be used for malicious purposes spanning from junk news and fake news or rumor spreading~\cite{sharma2019combating}, to propaganda and astroturfing \cite{Bovet2009, howard2017junk, neudert2017junk, howard2017junk_fr}. Specifically, an application is developed in~\cite{hui2020botslayer} to track information spreading on Twitter and tweets and accounts associated with suspicious campaigns. 

Usually, a legitimate bots' usage is adopted, to perform automated communication or administration during the electoral periods~\cite{howard2018algorithms}. However, Twitter bots have been extensively used for opinion hijacking during the Russian elections~\cite{krebs2011twitter, shane2017fake, lightfoot2017political, illing2018cambridge}, the 2017 French presidential election~\cite{Ferrara_2017}, the US elections~\cite{howard2018algorithms, byrnes2016bot, rizoiu2018debatenight}, the Catalan independence referendum~\cite{stella2018bots}, as well as in the Australian~\cite{waugh2013influence}, the Ukrainian ~\cite{hegelich2016social} and the Brazilian electoral process~\cite{arnaudo2017computational}. 
In~\cite{luceri2019evolution}, the authors study $\num[group-separator={,}]{245000}$  accounts on Twitter during the US 2016 presidential election and 2018 midterm elections, and they detect approximately $\num[group-separator={,}]{31000}$  bots. Forty-three million elections-related tweets of ongoing U.S. Congress investigation of Russian interference in the 2016 U.S. election campaigns are examined in~\cite{badawy2018analyzing}, where it is estimated that 4.9\% and 6.2\% of liberal and conservative users, respectively, were bots, with reported precision and recall scores above 90\%. In~\cite{doi:10.1080/10584609.2018.1526238}, the authors provide an analysis of the German parties' posts on Twitter from before and during the 2017 electoral period, and they reveal an increased amount of social bots (7.1\%  to 9.9\%). Other studies focusing on Twitter bots analysis include~\cite{stukal2017detecting} studying a specific consequential period in Russian politics (February 2014 to December 2015) and apply sentiment analysis or attempt to predict the results of the elections~\cite{7395825, antonakaki2017social}.


The authors in~\cite{Garimella2017ICWSM} investigate the political polarization on Twitter between 2009 and 2016, with an increased polarization of $10\%$ and $20\%$ being reported. The impact of Twitter bots during the first U.S. presidential debate of 2016 is studied in~\cite{rizoiu2018debatenight}, where a novel algorithm for estimating user influence from retweet cascades is introduced towards analyzing the role and user influence of bots versus humans. Moreover, a Twitter data analysis has been conducted in~\cite{Fraisier2018ICWSM} related to the 2017 French presidential campaign. The authors built a large and complex dataset of $\num[group-separator={,}]{22853}$ active Twitter profiles during the campaign from November 2016 to May 2017. Analysis of political discourse on Twitter in elections dataset has been noted during the US 2016 presidential elections \cite{YAQUB2017613}. Opinion hijacking has been observed not only in politics, but also in anti-vaccination promotion movements~\cite{broniatowski2018weaponized}. Thus, it is important to quantify the spread of fake news on Twitter~\cite{waugh2013influence} and the inherent variability~\cite{vosoughi2018spread}, in order to distinguish bots from human agents and legitimate users~\cite{EDWARDS2014372}.


It is evident that Twitter bot detection is a complex task, often requiring rigorous and solid treatment. Several ML-based solutions have been proposed. In particular, a real-time detection system dubbed as BotOrNot using a total amount of $\num[group-separator={,}]{1200}$ different features in combination with a Random Forest classifier is introduced in~\cite{10.1145/2872518.2889302}. An updated version of this system is described in~\cite{yang2019arming} named as Botometer, which requires Twitter API keys to collect user information during the real-time computations, thus it is not efficient to use real-time labeling tools in the case of big datasets. BotSentinel~\cite{BotSentinel} on the other hand, is a non-real-time labeling tool, capable of processing large amounts of user accounts and storing the results in a database. 
BotSentinel's offline labeling methodology is adopted whenever a user account is being suspended or removed, whereas real-time labeling does not provide any suspended account information. Moreover, the offline implementation allows us to increase the query rate limits, since it involves no labeling computational costs. 

Since a fundamental part of the bot detection pipeline corresponds to the computation of features based on Twitter data, a plethora of different types of features have been proposed. Various features are based on content~\cite{ahmed2013generic, gilani2017classification, lee2010uncovering, 10.1145/2872518.2889302, varol2017online}, sentiment~\cite{loyola2019contrast, 6921650, ferrara2016rise, loyola2019contrast}, account information~\cite{wald2013predicting, chu2012detecting, 10.1145/2872518.2889302, lee2010uncovering, loyola2019contrast}, usage~\cite{chu2012detecting} and network characteristics \cite{feng2020towards, keller2017manipulate, cresci2017paradigm}.

There is a growing number of ML and data (statistical) analysis-based Twitter bot identification tools. The most popular can be considered the
Stweeler tool~\cite{10.1145/2872518.2889360}, the Debot system~\cite{chavoshi2016identifying}, which takes into account synchronous bots spreading content, the TSD Sybil Detector~\cite{alsaleh2014tsd} that adopts a ML approach using 17 Twitter data-based features and the Retweet-Buster (RTbust)~\cite{mazza2019rtbust} which is an unsupervised learning tool combining feature extraction and clustering techniques. Moreover, sentiment analysis has been incorporated into the bot detection pipeline~\cite{6921650, loyola2019contrast}. A set of sentiment features is also exploited by the BotOrNot tool in~\cite{varol2017online}. The promising direction of ML-based Twitter bot detection can be reflected in DARPA competition, where six different research groups competed in performing bot identification, using anti-vaccination campaigns Twitter data~\cite{Subrahmanian_2016}.

\section{Methodology}\label{sec:methodology}

\subsection{Dataset}\label{sec:dataset}

\begin{table}[tp]
\begin{tabular}{|c||c|}
\hline
\textbf{Hashtag} & \textbf{Tweet Counts} \\
\hline\hline
\multicolumn{1}{|l||}{\texttt{\#VOTE}} & $\num[group-separator={,}]{3064099}$ \\
\multicolumn{1}{|l||}{\texttt{\#Trump202}} & $\num[group-separator={,}]{2403586}$ \\
\multicolumn{1}{|l||}{\texttt{\#Vote}} & $\num[group-separator={,}]{2200954}$ \\
\multicolumn{1}{|l||}{\texttt{\#Election2020}} & $\num[group-separator={,}]{1906959}$ \\
\multicolumn{1}{|l||}{\texttt{\#vote}} & $\num[group-separator={,}]{1838645}$ \\
\multicolumn{1}{|l||}{\texttt{\#Biden}} & $\num[group-separator={,}]{1\,063\,265}$ \\
\multicolumn{1}{|l||}{\texttt{\#Debate2020}} & $\num[group-separator={,}]{839717}$ \\
\multicolumn{1}{|l||}{\texttt{\#BidenHarris2020}} & $\num[group-separator={,}]{781697}$ \\
\multicolumn{1}{|l||}{\texttt{\#VoteBlueToSaveAmerica}} & $\num[group-separator={,}]{746896}$ \\
\multicolumn{1}{|l||}{\texttt{\#Trump}} & $\num[group-separator={,}]{601516}$ \\
\hline
\end{tabular}
\caption{Most popular HTs in our dataset. Tweets may contain multiple HTs so that the sum of tweets is not equal to the number of tweets in our collection.}
\label{table:hashtags}
\end{table}

In order to capture the US 2020 elections' Twitter dynamics shortly before the elections day (November 3rd 2020), we build a dataset where the most popular hashtags (HTs) related to the US 2020 elections were initially obtained.
We use Twitter API to retrieve all the tweets containing these HTs, spanning from September 1st, 2020 to November 3rd, 2020, resulting in a dataset of $15.6$ million tweets and $3.2$ million users. The ten most popular HTs are shown in Table~\ref{table:hashtags}. 


\subsection{Twitter Users Labeling}\label{sec:labeling}
The acquired dataset does not contain explicit knowledge whether a user is a bot or not. Since the goal of the current study is to provide a supervised ML-based solution for Twitter bot detection, it is crucial to obtain a bot vs. normal users labeled dataset.
Unfortunately, in the area of Twitter bot detection, it is not possible to collect accurate ground truth labels, without using third-party bot labeling tools. The classic solution of ground truth generation corresponds to a manual/crowd-sourcing analysis, which requires a thorough inspection of Twitter accounts, by human experts to identify the label of each account (via a majority voting rule). The manual labeling process is cumbersome when considering both the size of big datasets that contain millions of users (in our case the dataset contains $3.2$ million users) and the sophistication level of bot accounts which has risen during the last years.

As a means of overcoming the inherent restrictions of manual labeling, we utilize off-the-shelf ML-based techniques, allowing us to scale up the labeling procedure. ML methods achieve higher accuracy, in terms of ground truth labeling, as compared with the manual/crowd-sourcing analysis, since they exploit Twitter data feature representations not evident to human experts. Here, we use the Botometer~\cite{botometerOnline,varol2017online} and BotSentinel~\cite{BotSentinel} online tools to obtain the user labeling information. To achieve highly confident results, we combine the set of labels provided as output by the Botometer and the BotSentinel tool, respectively. In particular, we compute the intersection of the two label sets. The intersection contains the labels that are equal in both label sets. The users identified as bots by one of the two tools are marked as unlabeled.

Both bot detection labeling tools yield an output score for each requested Twitter account. The Botometer score lies in the interval $[0,5]$, while the BotSentinel score takes integer values in $\{ 0,\ldots,100 \}$. The higher the output score is, the higher the probability the requested account is a bot. A Twitter user is labeled as bot when the Botomoter and BotSentinel's output score is greater than $4.0$ and $75$, respectively. When the Botomoter and BotSentinel's output score is less than $1.0$ and $25$, respectively, the Twitter user is labeled as normal.
As mentioned above, since none of the two tools guarantees 100\% bot identification  accuracy, we aim at combining the scores from both tools and take into consideration the labels that are (mutually) equal. When an account is already suspended, Botometer cannot query Twitter API, and thus  
we perform a Twitter API check to identify whether an account is suspended or not.

\begin{table}[h]\centering
\begin{tabular}{ | c || c | c | c |}
  \hline
   \textbf{Step} & \textbf{Bot Users} & \textbf{Normal users} & \textbf{Total}\\\hline\hline
  Before labeling & \_ & \_ & 1.3$M$ \\ \hline
  BotSentinel & $\num[group-separator={,},group-minimum-digits={3}]{10324}$ & $\num[group-separator={,},group-minimum-digits={3}]{25546}$ & $\num[group-separator={,},group-minimum-digits={3}]{35870}$ \\ \hline
  Botometer & $\num[group-separator={,},group-minimum-digits={3}]{2180}$ & $\num[group-separator={,},group-minimum-digits={3}]{7267}$ & $\num[group-separator={,},group-minimum-digits={3}]{9447}$ \\ \hline
  Suspended & $\num[group-separator={,},group-minimum-digits={3}]{2389}$ & 0 & $\num[group-separator={,},group-minimum-digits={3}]{2389}$ \\ \hline \hline
  Final & $\num[group-separator={,},group-minimum-digits={3}]{4569}$ & $\num[group-separator={,},group-minimum-digits={3}]{7267}$ & $\num[group-separator={,},group-minimum-digits={3}]{11836}$ \\ \hline
  \hline
\end{tabular}
\caption{The number of users during each phase of the labeling of our dataset (the symbol $M$ corresponds to million).}
\label{table:user_each_phase}
\end{table}


\begin{figure}[t]
\centering
\includegraphics[width=80mm,scale=1.5]{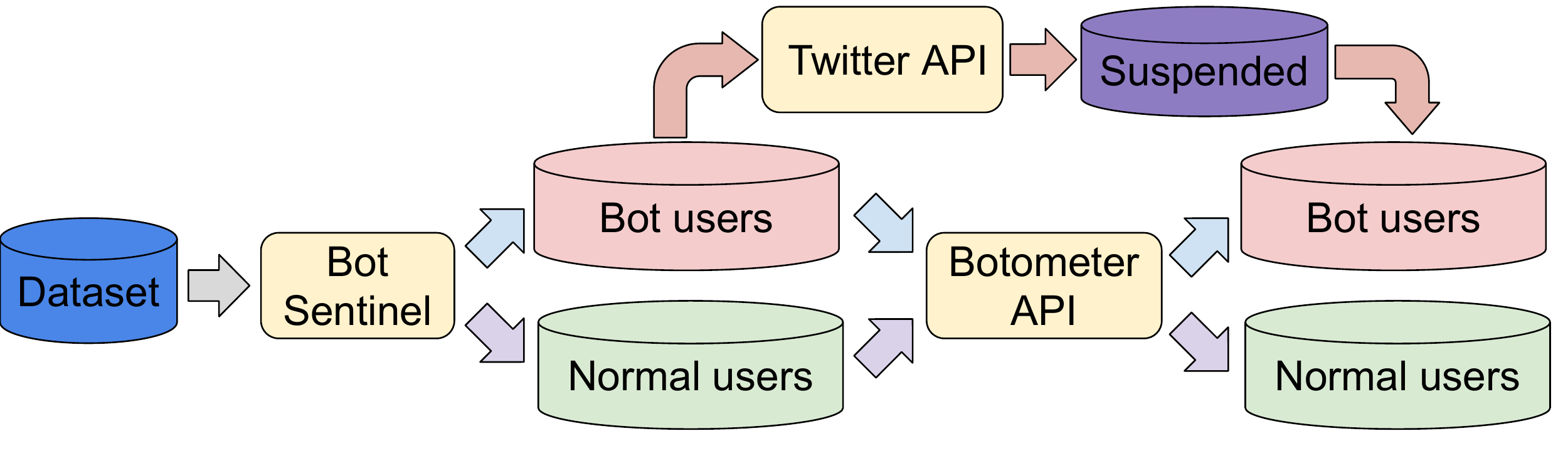}
\caption{The bot vs. normal users labeling pipeline.}
\label{labelPipe}
\end{figure}

To minimize the time complexity of the labeling process, we separate our dataset into two parts. The first part contains data extracted during September 2020 and is utilized for user labeling, ML model fine-tuning, training, validation and testing purposes. The second part incorporates data from October 1st, 2020 until November 3rd, 2020 and is used to evaluate the generalization capability of the proposed ML-based Twitter bot identification system on unseen data. 

\begin{table*}[ht]
\caption{Profile features extracted from Twitter user objects.}  \label{table:userFeatures}
\centering
\begin{tabular}{ | c  c ||c  c  c |}
  \hline 
  \textbf{Feature} & \textbf{Type} & \textbf{Feature} & \textbf{Type} & \textbf{Calculation}\\ \hline\hline
  \multicolumn{1}{|l}{\textit{statuses\_count}}  & count & \multicolumn{1}{|l}{\textit{screen\_name\_len}} & count&  \\ \hline
  \multicolumn{1}{|l}{\textbf{entities\_count}} &  count & \multicolumn{1}{|l}{\textit{description\_len}} & count & \\ \hline
  \multicolumn{1}{|l}{\textit{followers\_count}} & count  & \multicolumn{1}{|l}{\textit{screen\_name\_likelihood}} &  real-valued & \multicolumn{1}{l|}{likelihood of screen\_name}  \\ \hline
  \multicolumn{1}{|l}{\textit{friends\_count}} & count & \multicolumn{1}{|l}{\textbf{name\_screen\_sim}} & real-valued & \multicolumn{1}{l|}{name and screen\_name similarity}\\ \hline
  \multicolumn{1}{|l}{\textit{favourites\_count}} & count  &  \multicolumn{1}{|l}{\textbf{tweet\_retweet\_ratio}} & real-valued &  \multicolumn{1}{l|}{statuses\_count / retweet\_count score} \\ \hline
  \multicolumn{1}{|l}{\textit{listed\_count}} & count  & \multicolumn{1}{|l}{\textit{name\_digits}} & real-valued & \multicolumn{1}{l|}{number of digits in user name} \\ \hline
  \multicolumn{1}{|l}{\textit{name\_len}} & count  & \multicolumn{1}{|l}{\textit{screen\_name\_digits}} & real-valued & number of digits in user screen\_name \\ \hline
  \multicolumn{1}{|l}{\textbf{geolocation}} & boolean & \multicolumn{1}{|l}{\textit{tweets\_by\_age}} & real-valued & \multicolumn{1}{l|}{statuses\_count / user age} \\ \hline
  \multicolumn{1}{|l}{\textbf{protected}} & boolean &  \multicolumn{1}{|l}{\textit{followers\_by\_age}} & real-valued & \multicolumn{1}{l|}{followers\_count / user age} \\ \hline
  \multicolumn{1}{|l}{\textbf{location}} &  boolean &  \multicolumn{1}{|l}{\textit{friends\_by\_age}} & real-valued & \multicolumn{1}{l|}{friends\_count / user age} \\ \hline
  \multicolumn{1}{|l}{\textit{background\_img}} &  boolean  & \multicolumn{1}{|l}{\textit{favourites\_by\_age}} & real-valued & \multicolumn{1}{l|}{favourites\_count / user age} \\ \hline
  \multicolumn{1}{|l}{\textit{default\_profile}} &  boolean  & \multicolumn{1}{|l}{\textit{listed\_by\_age}} & real-valued & \multicolumn{1}{l|}{listed\_count / user age} \\ \hline
  \multicolumn{1}{|l}{\textit{verified}} &  boolean &  \multicolumn{1}{|l}{\textit{followers\_friends}} & real-valued & \multicolumn{1}{l|}{followers\_count / friends\_count} \\ \hline 
  \hline
\end{tabular}
\end{table*}

The dataset separation allows us to reduce the labeling process (computational) time without significant information loss, since the accounts remain active throughout the whole period of September and October. The first part has $1.3$ million users and more than $5$ million tweets and retweets, while the second part consists of $2.6$ million users and $10.6$ million tweets and retweets. A subset of users remain active during both periods, therefore it is obvious to notice the overlap between the two parts.


Figure~\ref{labelPipe} shows the labeling pipeline. Our dataset has $1.3$ million users during the BotSentinel labeling step. Then, the Botometer tool receives as input $\num[group-separator={,}]{35870}$ users, i.e.,  $\num[group-separator={,}]{10324}$ bot users and $\num[group-separator={,}]{25546}$ normal users (see Table~\ref{table:user_each_phase}), 
and outputs $\num[group-separator={,},group-minimum-digits={3}]{9447}$ users ($\num[group-separator={,},group-minimum-digits={3}]{2180}$ Twitter accounts labeled as bots and $\num[group-separator={,},group-minimum-digits={3}]{7267}$ Twitter accounts marked as normal accounts). As a parallel step, we query the Twitter API and the response provides a set of $\num[group-separator={,},group-minimum-digits={3}]{2389}$ users labeled as suspended. Therefore, the final labeled set has $\num[group-separator={,}, group-minimum-digits={3}]{4569}$ bot users and $\num[group-separator={,},group-minimum-digits={3}]{7267}$ normal users. Note that the overall labeling procedure is initialized with BotSentinel, since it does not impose any daily query limitations, in contrast with the Botometer. In the case of a Botomoter-based initialization step, the labeling outcome of the pipeline depicted in Figure~\ref{labelPipe} will be the same, but the processing time will grow dramatically and will require $650$ days to terminate, due to the Botometer request limitations. So, starting with BotSentinel will require only $18$ days.

We already mentioned that none of the existing labeling tools provide 100\% accurate ground truth labels. To quantify the accuracy of our proposed labeling pipeline, we compare the labeling results of our pipeline against the Twitter bot detection algorithm after a period of six months. According to Twitter, the number of bot accounts is $\num[group-separator={,},group-minimum-digits={3}]{4569}$ (with 51\% and 34\% of the bot accounts being suspended and removed, respectively), while $\num[group-separator={,},group-minimum-digits={3}]{7267}$ Twitter accounts are identified as normal (with only 1.9\% and 6.3\% of the normal accounts being suspended and removed, respectively). Twitter's labeling mechanism incorporates a lag time, and thus it cannot be efficiently used to compute our ground truth labels. Specifically, we manually confirmed that the lag time corresponds to about two months in the case of the US 2020 Elections.

\subsection{Feature Extraction}\label{section:FS}
Twitter API allows the collection of tweets, including information such as tweet text, tweet post time, as well as metadata such as HTs, URLs, and mentions. In this paper, we also include the user profile information by retrieving user objects, where all different types of the retrieved Twitter content are utilized, leading to a total amount of $335$ computed features. The features can be divided into four categories, namely, user profile, user context, user time, and user interaction.

\subsubsection{Profile Features}\label{section:User_features}
Twitter API retrieves user objects containing critical information to achieve accurate bot identification performance. The importance of user profile features is analyzed in various works~\cite{chu2012detecting,wald2013predicting,gilani2017classification,scalable2020aaai}. Typically, a user profile object includes user description, username, profile picture, and profile statistics (e.g., number of followers, friends, favourites, and listed). In this paper, bot vs. normal users distinction is promoted, by enriching the features set through the
extraction of profile features. For this, the user profile description and the user/screen name digits are taken into consideration.
The computed user object-based features correspond to unedited parameters such as the number of followers, friends, favourites, listed lists, and description length. Flag type elements like location usage, account description, protected flag, geolocation usage, and background image usage, are also estimated. Additional parameters such as the Jaccard similarity of the user and the account screen name are pre-computed and included in the overall features set. Table~\ref{table:userFeatures} shows the list of the extracted feature set, where the feature names written in italics correspond to the statistical features described in~\cite{scalable2020aaai}, and the feature names written in bold correspond to our proposed features leading to a $26$-dimensional profile feature space.

\begin{table*}[tp]\centering
\begin{tabular}{ | l || l |}
  \hline 
  \textbf{Feature} & \textbf{Description}  \\ \hline\hline

  \textit{N\_tweet\_mentioned\_tfidf} & TF-IDF score of the 3 most popular user mentions contained in tweets \\ \hline
  \textit{N\_tweet\_mentioned\_word} & The 3 most popular mentions in user tweets as word features \\ \hline
  \textit{N\_tweet\_hashtags\_tfidf} & TF-IDF score of the 3 most popular user HTs contained in tweets \\ \hline
  \textit{N\_tweet\_hashtags\_word} & The 3 most popular HTs in user tweets as word features \\ \hline
  \textit{N\_retweet\_mentioned\_tfidf} & TF-IDF score of the 3 most popular user mentions contained in RTs \\ \hline
  \textit{N\_retweet\_mentioned\_word} & The 3 most popular mentions in user RTs as word features \\ \hline
  \textit{N\_retweet\_hashtags\_tfidf} & TF-IDF score of the 3 most popular user HTs contained in RTs \\ \hline
  \textit{N\_retweet\_hashtags\_word} & The 3 most popular HTs in user RTs as word features \\ \hline
  \textit{N\_tweet\_word} & The 3 most popular words used by user in tweets as word features \\ \hline
  \textit{N\_retweet\_word} & The 3 most popular words used by user in RTs as word feature \\ \hline
  \textit{tweet\_number\_of\_urls} & Number of URLs in tweets, computed as average and standard deviation \\ \hline
  \textit{retweet\_number\_of\_urls} & Number of URLs in RTs, computed as average and standard deviation \\ \hline
  \textit{tweet\_number\_of\_hashtags} & Number of HTs in tweets, computed as average and standard deviation \\ \hline
  \textit{retweet\_number\_of\_hashtags} & Number of HTs in RTs, computed as average and standard deviation \\ \hline
  \textit{tweet\_number\_of\_mentions} & Number of mentions in tweets, computed as average and standard deviation \\ \hline
  \textit{retweet\_number\_of\_mentions} & Number of mentions in RTs, computed as average and standard deviation \\ \hline

  \hline
\end{tabular}
\caption{Context features based on user tweets and RTs, crawled by Twitter API.}
  \label{table:contextFeatures}
 \end{table*}

\subsubsection{Context Features}
The user profile feature set described in the previous section reflects
the statistics of the user's Twitter account, since the first day of the subscription to Twitter. However, the profile features lack semantic information regarding the actual content sent by the user. Thus, it is essential to incorporate contextual information such as user posts' content, most important user tweeted/re-tweeted topics, popular user HTs and number of URLs usage per tweet.
Table~\ref{table:contextFeatures} summarizes the list of the estimated context features.

Tweet's context characteristics provide a diverse range of uniqueness because each user operates in a different form of expression. To estimate the most frequent words and entities, we compute a subset of the context features such as the three most popular words, mentions and HTs per user (punctuation marks and stop words are removed since they do not provide important information). For each user, we discover the most frequent sentences (user mentions, hashtags, upper and lower words). User mentions and hashtags may provide unique information that highlights the characteristics of a particular user, thus we compute the term frequency-inverse document frequency (TF-IDF)~\cite{rajaraman_ullman_2011} on the collected dataset. This allows us to identify the importance level of the user's hashtags and mentions. In particular, we compute the TF-IDF of the overall user mentions and hashtags and for each particular user, we identify the three most frequent mentions/hashtags. The final step is to compute the TF-IDF based on the overall frequency.

\begin{table}[b]\centering
\begin{tabular}{|p{0.3\textwidth} || p{0.5\textwidth}|}
  \hline 
  \textbf{Feature} & \textbf{Description}  \\ \hline\hline
  \textit{daily\_rt} &  RT \%  each week day \\ \hline
  \textit{daily\_tw} & tweets \% each week day \\ \hline
  \textit{daily\_rt\_tw} &  tweets/RT \% each week day \\ \hline
  \textit{daily\_retweet\_avg} & average daily number of RTs\\ \hline
  \textit{daily\_tweet\_avg} & average daily number of tweets \\ \hline
  \textit{hourly\_rt} &  RTs \% of daily hours \\ \hline
  \textit{hourly\_tw} &  tweets \% of daily hours \\ \hline
  \textit{hourly\_rt\_tw} & tweets/RTs \% of daily hours \\ \hline  
  \textit{retweet\_time} & time difference between \\  & original tweet and user RT, computed as min/max/avg/std \\ \hline
\end{tabular}
\caption{Time-based features computed on user's tweet/RT object. Min, max, avg and std correspond to minimum, maximum, average and standard deviation, respectively.}
  \label{table:timeFeatures}
 \end{table}

The next step is to use the word2vec algorithm~\cite{wordTovec} to learn the word embeddings from the obtained Twitter dataset, allowing us to transform text-based features into a $10$-dimensional space. The most frequent words, mentions and HTs are transformed with the trained word2vec model. Note that the text-based features might differ between the user's original tweets and RTs, since they are usually written by a different user. Thus, text-based features are computed separately for each user's tweets and RTs.

\subsubsection{Time-Based Features}

The automated bot accounts follow a non-uniform time distribution activity~\cite{zhang2011detecting}, either due to Twitter API time constraints regarding tweet posts within short time intervals, or as a result of the job schedulers that invoke tasks at specific time intervals. In addition, the automated bots follow a non-uniform activity pattern whenever scripts are scheduled to start or stop running at the same timestamps. Thus, the automated bots behaviour can be detected by recognizing extremely non-uniform or highly uniform tweet posts time patterns. On the other hand, normal users' tweets typically follow a diurnal pattern, which can be predictable for specific users. As mentioned in~\cite{chu2010tweeting}, the human activity follows a special pattern on Twitter since humans perform tweet posts at specific daily time intervals, while the activity appears to be lower during the weekends. Nevertheless, bots' activity pattern is more unpredictable because it does not follow the same activity level per day. The automated behaviour of a bot accounts is constantly evolving,  almost mimicking human users activity, making it a challenging task to detect bot accounts solely based on time-oriented features. However, due to the fact that not all automated behaviours are similar, we aim to compute and use time-based patterns as additional input to the proposed ML pipeline to enhance the bot vs. normal users detection accuracy.

In this paper, we extract multiple time-based features such as the RT time, as well as the hourly and daily activity. Regarding the RT times, we compute the difference between the original tweet and the RT time provided in the tweet object. We also measure the RT time distribution per user, where the minimum, maximum, average and standard deviation values of the RT time are included in the feature set. As an account activity metric, the daily percentage of tweets and RTs is computed (i.e., we can identify during which days the users appear to be more active). Similar metrics are estimated during the active days and hours, and thus we can identify the exact hourly intervals of the day in which the user is vigorously posting tweets or RTs. Table~\ref{table:timeFeatures} presents the set of time-based features.

\begin{figure}[tp]
\centering
\includegraphics[width=\linewidth]{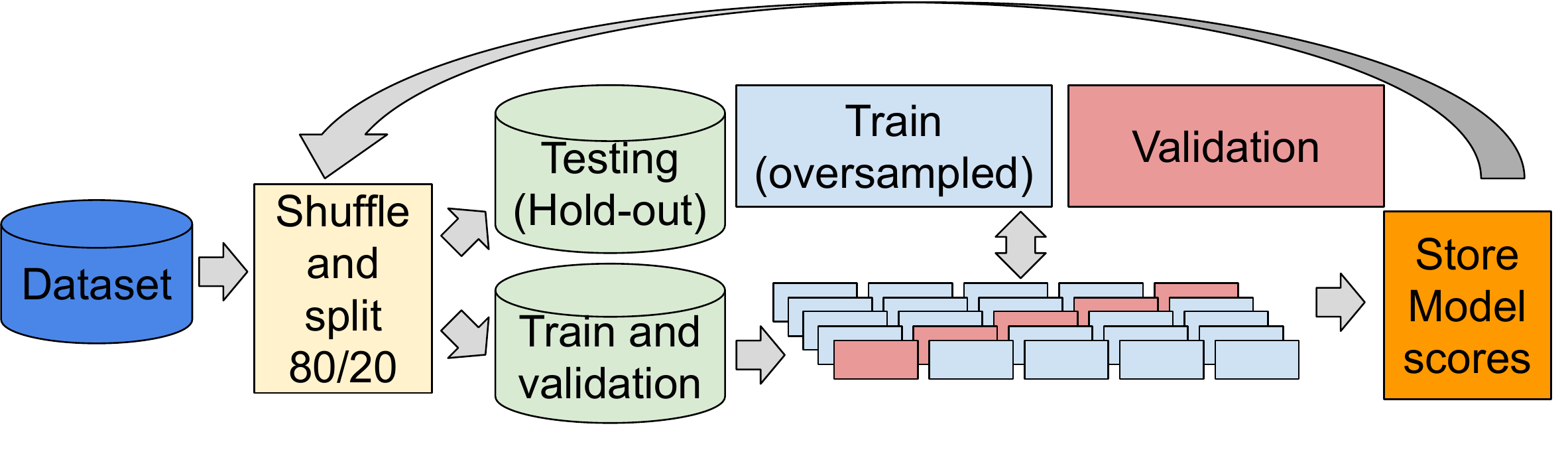}
\caption{ML model selection pipeline.}
\label{fig:pipeline}
\end{figure}

\subsubsection{Interaction Features}
The final set of extracted features is based on the RT network graph which models user interactions.
The RT graph is estimated based on the collected dataset, with the nodes representing users and the directed edges defining a RT action from user $i$ to user $j$.
The edge weight indicates the number of RTs between the two users. The resulted graph represents the network of the RT connections in our dataset. Finally, we use Gephi~\cite{ICWSM09154} to compute the node statistics such as in-degree and out-degree of each node. 

\section{Experimental Results}\label{sec:results}
In this section, we examine the performance of our proposed system, with respect to the resulting bot vs. normal users detection accuracy.

\subsection{ML Framework}\label{section:ML_framework}
 

As a main step towards building a robust and accurate ML-based bot identification system, we perform a model selection procedure by examining the bot vs. normal users classification accuracy of several state-of-the-art ML algorithms. In particular, we evaluate the performance of Random Forest~\cite{breiman2001random}, Support Vector Machine (SVM)~\cite{cortes1995support} and  Extreme  Gradient  Boosting (XGBoost)~\cite{XGBoost2016} algorithm. Each ML method involves a different number of hyper-parameters, and thus it is of paramount importance to follow a hyper-parameter tuning procedure to identify the best (trained) version of each ML model and promote a fair models' comparison. 
Figure~\ref{fig:pipeline} illustrates the ML model selection
pipeline based on a combination of $80/20$ train/test split (using random shuffle) and a 5-fold cross-validation scheme, i.e., the dataset is randomly shuffled, where $80\%$ of the dataset is used for training/validation and the rest $20\%$ (hold-out part) of the dataset is exploited for testing purposes. 
Each train/test split is performed in a stratified manner in order to have the same ratio of classes in both training and testing data. During the 5-fold cross-validation process, the synthetic minority oversampling technique (SMOTE) using Tomek links~\cite{SMOTE_Tomek} is applied on the training folds to balance the two (bot vs. normal) distributions by oversampling the minority (bot) class distribution. 

\begin{table}[htb]\centering
\begin{tabular}{ | c || c | c | c |}
  \hline
   \textbf{Model} & \textbf{F1} & \textbf{PR-AUC} & \textbf{ROC-AUC}\\\hline\hline
  XGBoost & $0.919$  & $0.967$ & $0.979$  \\ \hline
  Random Forest & $0.908$  & $0.955$ & $0.973$ \\ \hline
  SVM & $0.889$  & $0.941$ & $0.964$  \\
  \hline
\end{tabular}
  \caption{Testing accuracy during the model selection phase.}
  \label{table:modelScores}
 \end{table}
Additionally, we employ feature selection based on three different methods, namely, Lasso~\cite{Tibshirani94regressionshrinkage}, Random Forest feature selection and model feature importance.
Among the three feature selection methods, the model feature importance\footnote{\href{https://scikit-learn.org/stable/modules/generated/sklearn.feature_selection.SelectFromModel.html}{Scikit-learn function: SelectFromModel}} provided the features having the highest predictive accuracy. 

It is important to mention that some of the context features are vectors provided by the pre-trained word2vec model. 
The word2vec model provides a 10-dimensional space representation of the text, but only a few out of the ten dimensions are informative for the model. For this, we keep only the informative dimensions through the feature selection process. An example is presented in Figure~\ref{fig:shap}, where the feature ``\textit{N1\_retweet\_hashtag\_word\_7}'' represents the first most popular hashtag seen in the user retweets. According to the feature name, the seventh element of this particular word (embedding) vector corresponds to the most informative dimension.
 
Table~\ref{table:modelScores} reports the F1 score and the two areas under the curve (AUC) scores, i.e., the precision-recall (PR) AUC value and the receiver operating characteristic (ROC) AUC value, averaged over ten repetitions. It can be seen that the XGBoost model achieves slightly better results on the testing data than SVM and Random Forest.
Thus, we select XGBoost as the basic ML model applied in the next experimental evaluation phase.

 \begin{table}[t]\centering
\begin{tabular}{ | l || r | r | }
  \hline
   \textbf{Dataset}  & \textbf{\# bots} & \textbf{\# normal} \\   \hline \hline 
   cavarlee & $\num[group-separator={,}]{15483}$  & $\num[group-separator={,}]{14833}$ \\ 
   varol-icwsm & $733$ & $\num[group-separator={,}, group-minimum-digits={3}]{1495}$ \\ 
   cesci-17 & $\num[group-separator={,}, group-minimum-digits={3}]{7049}$  & $\num[group-separator={,}, group-minimum-digits={3}]{2764}$\\ 
   pronbots & $\num[group-separator={,}, group-minimum-digits={3}]{17882}$  & $0$\\
   celebrity & $0$  & $\num[group-separator={,}, group-minimum-digits={3}]{5918}$ \\
   vendor-purchased & $\num[group-separator={,}, group-minimum-digits={3}]{1087}$  & $0$\\
   botometer-feedback & $139$  & $380$\\
   political-bots & $62$  & $0$\\
   Gilani-17 & $\num[group-separator={,}, group-minimum-digits={3}]{1090}$ & $\num[group-separator={,}, group-minimum-digits={3}]{1413}$\\
   Cresci-rtbust & $353$  & $340$\\
   cresci-stock & $\num[group-separator={,}, group-minimum-digits={3}]{7102}$   & $\num[group-separator={,}, group-minimum-digits={3}]{6174}$ \\
   Midterm-18 & $\num[group-separator={,}, group-minimum-digits={3}]{42446}$  & $\num[group-separator={,}, group-minimum-digits={3}]{8092}$ \\
   Botwiki & $698$  & $0$\\
   verified & $0$  & $\num[group-separator={,}, group-minimum-digits={3}]{1987}$ \\ \hline \hline
   \textbf{Total} & $\num[group-separator={,}, group-minimum-digits={3}]{94124}$  & $\num[group-separator={,}, group-minimum-digits={3}]{43396}$ \\
  \hline
\end{tabular}
   \caption{Publicly available labeled datasets used for bot detection performance evaluation in~\cite{scalable2020aaai}.}
  \label{table:aaaiDatasets}
 \end{table}

\subsection{General Model Comparison}
We evaluate the generalization capability of the XGBoost model, which is already fine-tuned on the US 2020 Elections dataset (see in section~\textit{\nameref{section:ML_framework}}), against a general model~\cite{scalable2020aaai} applied to detect bots on various datasets. To perform this comparison, we collect the public datasets provided by~\cite{scalable2020aaai} and we perform an experimental evaluation with similar specifications as described in~\textit{\nameref{section:ML_framework}} section. The authors in~\cite{scalable2020aaai} utilize only the statistical features computed on the user objects, without exploiting any further knowledge related with user interactions, RT times, or contextual information of user tweets. As a result, we extract and use the same feature set in both XGBoost and general model implementation to promote a fair comparison.

We follow the experimental strategy described in~\cite{scalable2020aaai}. Our proposed XGBoost model is trained over all possible combinations of publicly available Twitter datasets mentioned in Table~\ref{table:aaaiDatasets}. The dataset combinations that correspond to the best testing performance of each model are presented in Table~\ref{table:aaaiTable}.
The first six rows of Table~\ref{table:aaaiTable} indicate the different dataset combinations (check mark symbols) used as training data by the M196, M195, U1 and U2 models. The next four rows correspond to the identification accuracy of each model using as testing (unseen) data the Botwiki \& verified, Midterm-18, Gilani-17 and Cresci-rtbust datasets, respectively. Note that the ROC-AUC scores over the four datasets correspond to the M196 and M195 dataset combinations used in~\cite{scalable2020aaai} and to the U1, U2 dataset combinations trained by our model.
The inherent information of the various combined datasets reflect the differences between bot vs. normal users. This information can be learned via the XGBoost model, achieving robust generalization capabilities. 

The best ROC-AUC scores are achieved by our XGBoost model when using as training data the dataset combinations U1 (varol-icwsm, pronbots, botometer-feedback)  and U2 (varol-icwsm, pronbots, botometer-feedback, political-bots).
Table~\ref{table:aaaiTable} also reports the best ROC-AUC scores achieved by the Random Forest model in~\cite{scalable2020aaai} utilizing as training data the dataset combinations M196 (varol-icwsm, cesci-17, celebrity botometer-feedback, political-bots) and M195 (varol-icwsm, cesci-17, celebrity botometer-feedback).

\begin{table}[t]\centering
\begin{tabular}{ | c || c | c | c | c | }
  \hline
   \textbf{Dataset}  & \textbf{M196} & \textbf{M195} & \textbf{U1} & \textbf{U2}\\   \hline\hline
   varol-icwsm & \checkmark & \checkmark & \checkmark &  \checkmark\\ \hline
   cesci-17 & \checkmark & \checkmark &  & \\  \hline
   pronbots & & & \checkmark  & \checkmark \\ \hline
   celebrity & \checkmark & \checkmark &  & \\ \hline
   botometer-feedback & \checkmark & \checkmark & \checkmark & \checkmark \\ \hline
   political-bots & \checkmark & &  & \checkmark \\ \hline\hline
   Botwiki \& verified & $0.99$ & $0.99$ & $0.978$ & $0.978$ \\ \hline
   Midterm-18 & $0.99$ & $0.99$ & $0.954$ & $0.951$\\ \hline
   Gilani-17 & $0.68$ & $0.69$ & $0.75$ & $0.745$ \\ \hline
   Cresci-rtbust & $0.60$  & $0.59$ & $0.63$ & $0.614$\\ \hline \hline
   \textit{Average ROC-AUC}& \textit{0.815} &  \textit{0.815} & \textit{0.828} & \textit{0.822}\\ 
  \hline
\end{tabular}
   \caption{Training dataset combinations and performance results between our XGBoost model and the method described in~\cite{scalable2020aaai}.}
  \label{table:aaaiTable}
 \end{table}

The results clearly indicate that our fine-tuned models (U1 and U2) achieve superior results when compared with the general models (M195 and M196) in~\cite{scalable2020aaai}, as reflected on the average ROC-AUC scores.
According to the best performance (U1 model) across the different datasets, the ROC-AUC values range from $0.97$ to $0.63$. 
In order to identify how the U1 and U2 trained models will perform on the US 2020 Elections dataset, we measure the ROC-AUC score for these two models. The models U1 and U2 achieved $0.609$ and $0.618$ ROC-AUC, respectively.
For this purpose, we provide the performance of the model trained directly on the US 2020 Elections data.
 
\begin{table}[hb]\centering
\begin{tabular}{ | c || c | }
  \hline
   \textbf{Feature set}  & \textbf{Number of features} \\   \hline\hline       
   Statistical General only & 20 \\ \hline
   Statistical & 26 \\ \hline
   Context & 204 \\ \hline
   Time & 99 \\ \hline
   Graph & 6 \\ \hline
   Our model & 228 \\ \hline
\end{tabular}
   \caption{Number of features extracted by each feature category from the US 2020 Elections dataset.}
  \label{table:featuresByCat}
 \end{table}
 
\subsection{Statistical vs. Context Features}\label{section:statisticalVScontext}
\begin{figure}[tb]
    \centering
    \includegraphics[width=70mm,scale=1.5]{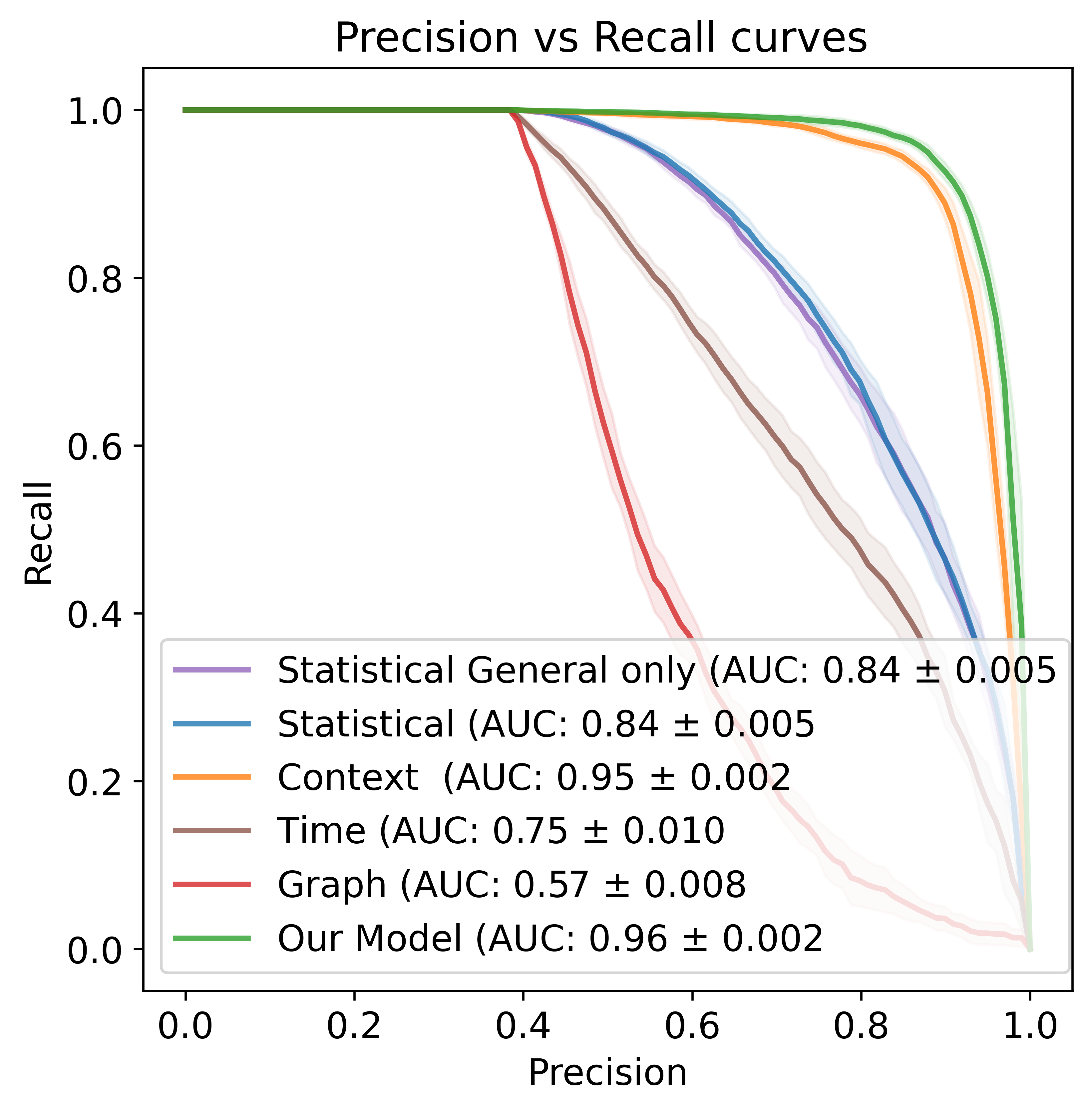}
    \caption{Mean PR curves: multiple types of features compared against our selected features.}
    \label{fig:precRecal}
\end{figure}

\begin{figure}[tb]
    \centering
    \includegraphics[width=70mm,scale=1.5]{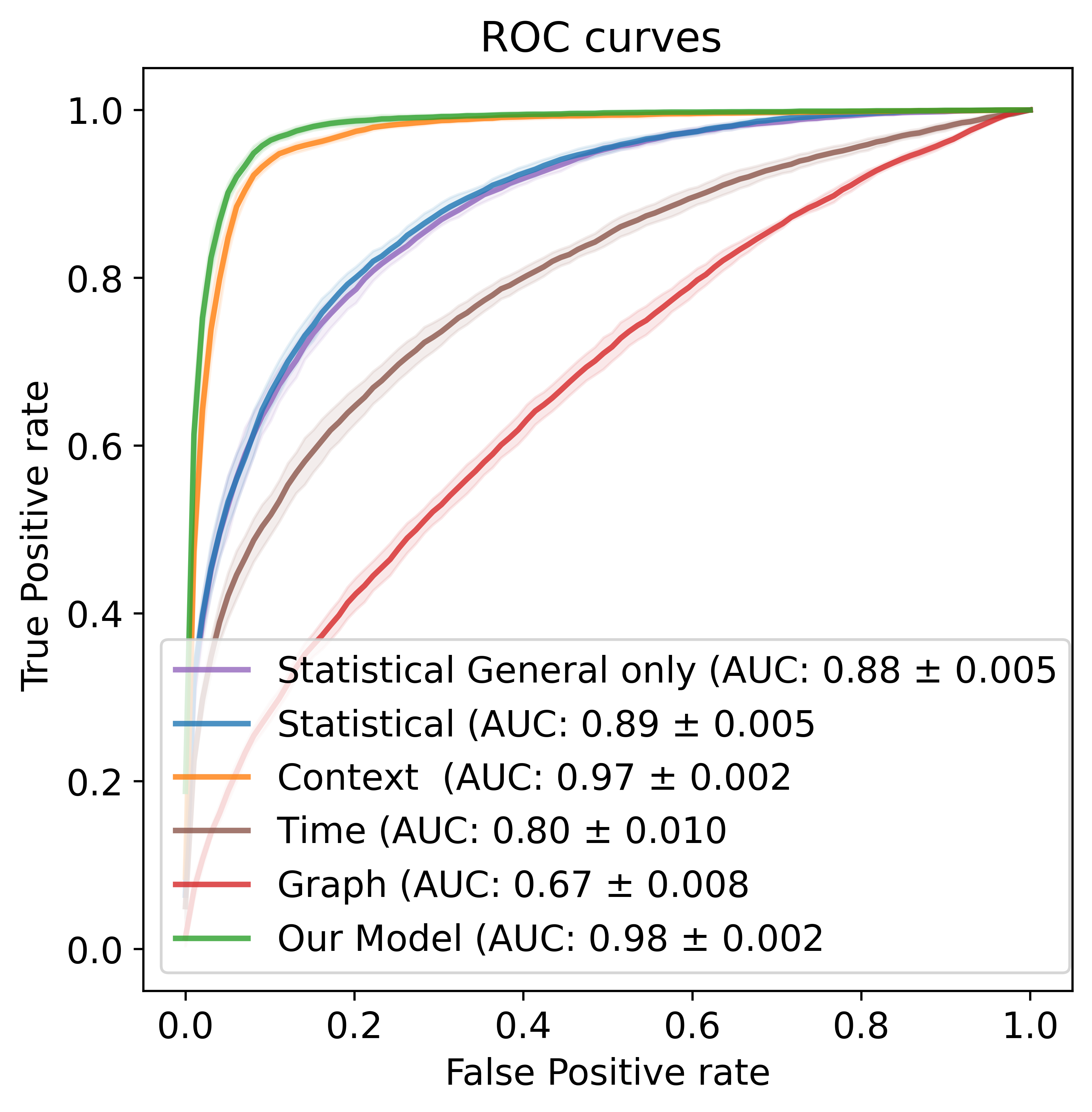}
    \caption{Mean ROC curves: multiple types of features  compared against our selected features.}
    \label{fig:roc}
\end{figure}

We compare the proposed XGBoost model with the model introduced in~\cite{scalable2020aaai} in light of the statistical features set. The authors in~\cite{scalable2020aaai} exploit only the statistical features set. To promote a fair models' comparison we use the statistical features alone, including the number of followers, listed, favorites, and friends, as well as the computation of the growth rate, based on the user account age, the number of digits in the screen name and the account screen name likelihood. The extracted set of features do not contain semantic information related with the posts content. The rest of the feature types described in section~\textit{\nameref{section:FS}} are utilized separately in our model, during the training and validation steps. Each feature category, presented in Table~\ref{table:featuresByCat} is used separately and compared to each category's performance against PR and ROC curves with the features described in~\cite{scalable2020aaai}, as well as the best features that were selected by our model.
Figure~\ref{fig:precRecal} presents the precision vs. recall performance of those features, with separate information of F1-score of the hold-out dataset portion. Figure~\ref{fig:roc} illustrates the ROC-AUC curve and the corresponding AUC score for each feature set.
According to the ROC-AUC performance model, utilizing a mixture of multiple features with proper feature selection results in a better ROC-AUC performance model, since each feature set contains critical information for the model. 



\subsection{Generalization Performance: US 2020 Elections Dataset}\label{sec:Performance}

The combination of multiple feature types provide the best bot identification accuracy as it is experimentally evaluated in section~\textit{\nameref{section:statisticalVScontext}}, where the number of multiple combined features is 228. We use this set of features to investigate the generalization capability of our XGBoost model on the US 2020 Elections dataset. In particular, we divide the US 2020 Elections dataset into two parts as mentioned in section~\textit{\nameref{sec:labeling}}, i.e., the first part corresponds to the time interval between September 1st and September 30th, while the second part corresponds to the interval between October 1st and November 3rd. The experimental specification is the same as that adopted in section~\textit{\nameref{section:ML_framework}}. The only difference is that the train/test split now is $70/30$, where $70\%$ of the September dataset is used for train/validation of the XGBoost model, while the rest $30\%$ is used for testing with the F1 score equal to $0.916$ and the ROC-AUC score is $0.98$. A difference between these results and the ones depicted in Figure~\ref{fig:precRecal} due to the random data shuffling.


The second dataset (October 1st to November 3rd) is also used as testing data, to evaluate the bot identification performance of the already trained (on the $70\%$ data of September) XGBoost model on unseen data that correspond to an extended time horizon. The XGBoost model achieves an average of $0.896$ F1 score and $0.977$ ROC-AUC.
The aforementioned results, clearly indicate that our proposed ML model achieves impressive generalization capabilities by identifying bot accounts on future data, based on past training samples.


\subsection{Model Explainability}\label{section:XAI}
One of the ultimate goals of the current paper is to ``unlock'' the
proposed ML model mechanism in order to better understand how the
model yields its predictions. We use SHapley Additive exPlanations
(SHAP) values proposed in~\cite{lundberg2017unified} since they
present several advantageous characteristics. First and most
importantly, SHAP values are model-agnostic, i.e., they are not bound
to any particular type of ML model. Secondly, SHAP values present
properties of local accuracy, consistency, and missingness, which are
not found simultaneously in other methods. Lastly, SHAP implementation
is actively supported by an open-source 
community\footnote{\url{https://shap.readthedocs.io/}}, it is well
documented and straightforward  to use.

Before proceeding to the SHAP values explanation, let us first, provide a description of the concept of Shapley value. More specifically, Shapley introduced a game-theoretic approach for assigning fair payouts to players depending on their contribution to the total gain~\cite{Shapley1953}.  Within a predictive modeling task, this translates to assigning an importance numerical value to features that depend on their contribution to a prediction. Thus, in the predictive ML context, a Shapley value can be defined as the average marginal contribution of a feature value across all possible feature coalitions. Based on this definition, a Shapley value for a given feature can be interpreted as the difference between the mean prediction for the whole dataset and the actual prediction.

The Shapley values are represented as a linear model of feature coalitions by the SHAP method~\cite{lundberg2017unified}. SHAP values exploit the game theory's Shapley interaction index, which allows allocating payouts, i.e., importance, not just to individual players, i.e., features, but also among all pairs of them. As a result, SHAP values can explain the modeling of local interaction effects, and allow the possibility of providing new insights into the ML model's features.

\begin{figure}[tbp]
    \centering
    \includegraphics[width=80mm,scale=1.5]{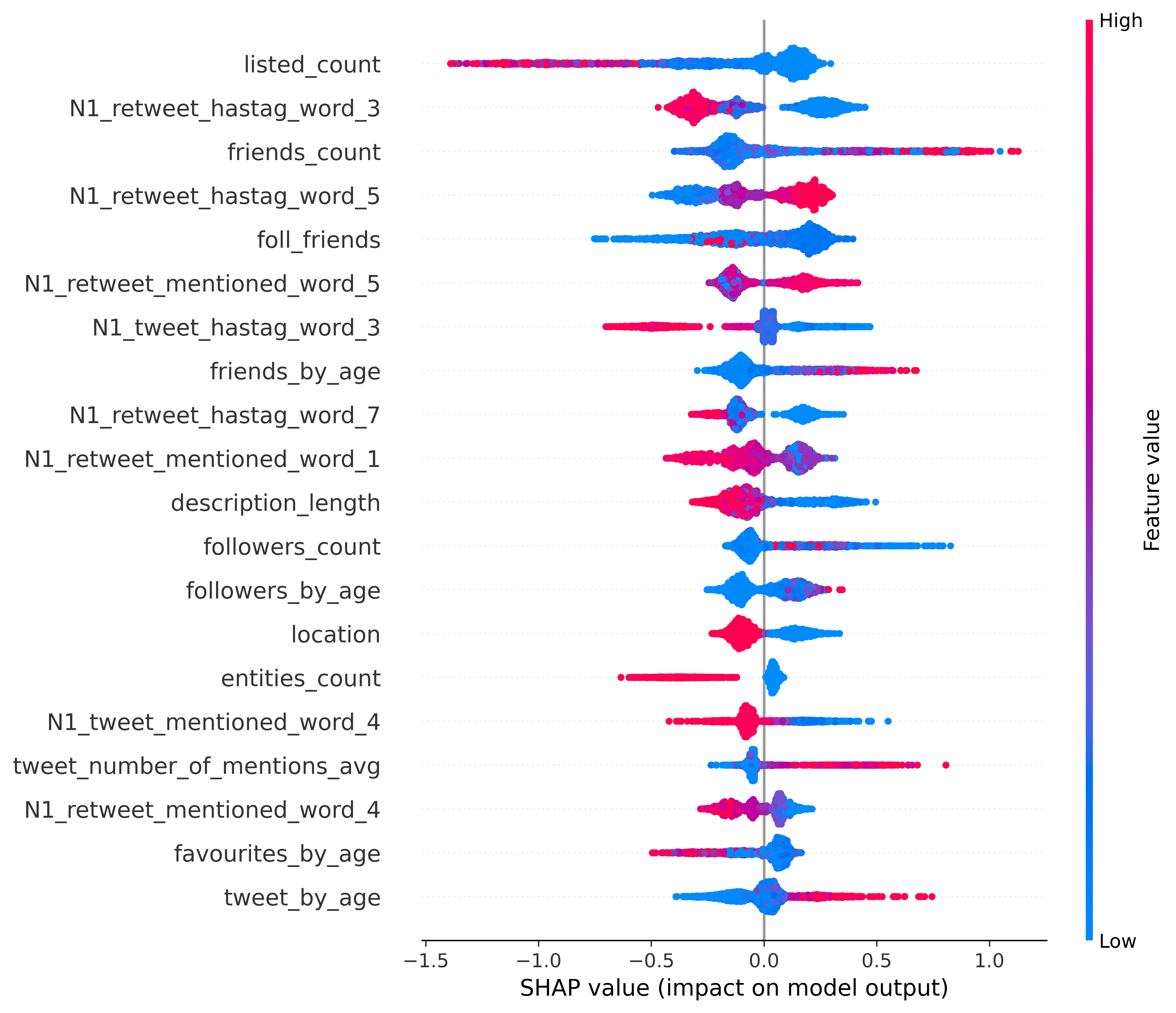}
    \caption{US 2020 Elections dataset: summary plot for SHAP values. The top twenty features with the highest impact at the XGBoost model's output are depicted.}
    \label{fig:shap}
\end{figure}

Figure~\ref{fig:shap} shows the summary plot for SHAP values related with the features extracted from the US 2020 Elections dataset. The top twenty features with the highest impact at the XGBoost model's output are depicted. For each feature, one point corresponds to a single Twitter user. A point's position along the $x$-axis (i.e., the actual SHAP value) represents the impact that feature had on the model's output for that specific Twitter user. Mathematically, this corresponds to the malicious behaviour risk relative across Twitter users (i.e., a Twitter user with a higher SHAP value has a higher risk being malicious relative to a Twitter user with a lower SHAP value). Features are arranged along the $y$-axis based on their importance, which is given by the mean of their absolute Shapley values. The higher the feature is positioned in the plot, the more important it is for the XGBoost model.

A further analysis of the results in Figure~\ref{fig:shap} indicates that the top twenty features with the highest impact on the XGBoost model's output correspond to statistical, time and graph-based features. 
In particular, features such as Twitter lists and average number of mentions in user tweets appear to have a high impact in XGBoost model's output. We expect that a combination of features with the highest output impact could provide the best possible bot identification performance. This statement can be confirmed by the results mentioned in section~\textit{\nameref{sec:Performance}}.  
Based on the SHAP values summary plot depicted in Figure~\ref{fig:shap}, it is obvious that ``\textit{listed\_count}'' corresponds to the feature with the highest impact at XGBoost model's bot vs. normal user detection. As shown in Figure~\ref{fig:shap} bot users tend to not belong to Twitter lists, whereas normal users could be members of more than one list. We can also deduce that bot users have lower values of ``\textit{favourites\_by\_age}'' (also known as likes), which means that bot users tend to ignore the like button of other users' posts. This could be explained by the complexity of bot account implementation. Finally, we notice that bot users have high values of ``\textit{friends\_by\_age}'' feature, which means that they tend to connect to more accounts within a short period of time. This activity is obvious since bot accounts try to gain high visibility and expand to larger parts of the Twitter network. Presented explanations confirm our initial intuitive explanations regarding the difference between normal and bot accounts activity.



\section{Conclusions and Future  Work}\label{sec:conclusion}
This paper introduces a novel methodology based on a supervised machine learning (ML) framework for identifying bot vs. normal Twitter users using a wide range of extracted features. Specifically, the proposed system incorporates the extraction and labeling of multiple features, with the ground truth labels estimated through the combination of two online bot detection tools' output. A thorough ML analysis involving train/validation/test split, feature selection, oversampling and hyper-parameters tuning, establishes the Extreme Gradient Boosting (XGBoost) algorithm as the best ML model along with a specific set of features. The selected XGBoost model when trained on a wide range of combined features spanning from profile and context features to time-based and interaction features achieves the highest bot detection accuracy.

The generalization capability of the proposed ML system is extensively examined through an experimental evaluation process, and compared with a recently introduced general model~\cite{scalable2020aaai}. Finally, the obtained explanations revealed meaningful insights from a Twitter data analysis point of view about the reasoning process behind the XGBoost model’s decisions. Future work concerns the extension of the proposed methodology by performing text analysis on the tweet corpus posted by the bot users, to identify the shared type of content during the US 2020 Elections period.

The code and the employed datasets are available at the following link in GitHub repository: \cite{github_twitterbots}

\section{Acknowledgements}
We would like to thank the reviewers for their valuable comments. This document is the result of the research projects CONCORDIA (grant number 830927), CyberSANE (grant number 833683) and PUZZLE (grant number 883540) co-funded by the European Commission, with  (EUROPEAN COMMISSION Directorate-General Communications Networks, Content and Technology).

\bibliography{paper}

\end{document}